\documentstyle[12pt]{article}
\newcommand{\be}{\begin{equation}}
\newcommand{\ee}{\end{equation}}
\newcommand{\R}{\rm I \mkern -3mu R}

\newcommand{\up}{\widetilde{{\cal P}^{\uparrow}_{+}}}
\newcommand{\ul}{\widetilde{{\cal L}^{\uparrow}_{+}}}

\newcommand{\xmp}{X_{m}^{+}}

\newcommand{\masp}{d\alpha^{+}_{m}}

\newcommand{\y}{Y_{m}}

\begin{document}
\thispagestyle{empty}
\begin{flushright}
IFA-FT-400-1994, October
\end{flushright}
\bigskip\bigskip\begin{center}
{\bf \Large{FREE FIELDS\\
{}~\\ FOR ANY SPIN IN RELATIVISTIC\\
{}~\\ QUANTUM MECHANICS}}
\end{center}
\vskip 1.0truecm
\centerline{\bf
D. R. Grigore\footnote{e-mail: grigore@roifa.bitnet, grigore@ifa.ro}}
\vskip5mm
\centerline{Dept. Theor. Phys., Inst. Atomic Phys.,}
\centerline{Bucharest-M\u agurele, P. O. Box MG 6, ROM\^ANIA}
\vskip 2cm
\bigskip \nopagebreak \begin{abstract}
\noindent
A general discussion of the construction of free fields based on
Weinberg anszatz is provided and various applications appearing
in the litterature are considered.
\end{abstract}

\newpage\setcounter{page}1

\section{Introduction}

In this letter we provide a general discussion of the procedure
of constructing a relativistic free quantum field when it is
required that the one-particle Hilbert space carries a certain
projective unitary representation of some invariance group. The
usual case is when the invariance group is the Poincar\'e group
and the representation is irreducible, corresponding to an
elementary system.

The discussion was initiated a long time ago by Weinberg \cite
{W1}-\cite{W4} who provided a rather exhaustive analysis of the
usual case when the fields are living in the four-dimensional
Minkowski space. Recently the analysis was extended to the
three-dimensional case with the purpose of obtaining free fields
for any spin \cite{G1}.

We will show that a general discussion of this problem is possible.
One recovers extremely easy the results alluded above and,
moreover one can treat other types of fields which describe
excitations which are not point-like. We have in mind, for
instance, fields localized on cones which appear in
three-dimensional quantum fields theory (see for instance \cite{M}.)

The general theory will be developped in Section 2 and the
applications in Section 3.

\section{Relativistic Quantum Field Theory}

2.1 We begin with a general discussion of the notion of
relativistic quantum field theory. As it is well known \cite{SW},
\cite{J}, \cite{BLT} a quantum field theory is an ansamble
$
({\cal H}, \Phi_{0}, D, \phi_{\alpha}(x) )
$
where
${\cal H}$
is a complex Hilbert space,
$
\Phi_{0}
$
is the vacuum state,
$D \in {\cal H}$
is a dense domain containing
$
\Phi_{0}
$
and
$\phi_{\alpha}(x)$
are some distribution-valued operators such that
$\phi_{\alpha}(x)$
and
$\phi_{\alpha}(x)^{*}$
are defined on
$D$
and are leaving it invariant. Here
$x$
takes values in some configuration manifold
$X$,
which is usually the Minkowski space. For a relativistic quantum
field one supposes moreover that an invariance group
$G_{inv}$
is given;
$G_{inv}$
acts on the manifold
$X$
and according to the basic principles of quantum theory {\cite
V} we also have an projective unitary representation of
$G_{inv}$
in
${\cal H}$.
We will consider the usual case when this representation follows
from a true representation
$
G \ni g \mapsto {\cal W}_{g} \in {\cal U}({\cal H})
$
of the universal covering group
$G$ of
$
G_{inv};
$
here
$
{\cal U}({\cal H})
$
is the set of the unitary operators in
${\cal H}$.
The general case can be also treated on similar lines.
The interpretation of these objects being obvious, one can argue
that the fields
$\phi_{\alpha}(x)$
should behave as follows:
\be
{\cal W}_{g} \phi_{\alpha}(x) {\cal W}_{g}^{-1} = c_{\alpha\beta}(g^{-1},
g\cdot x) \phi_{\beta}(g\cdot x)
\ee

We use the summation convention over the dummy indices. Here
$
c_{\alpha\beta}(g,x)
$
is a matrix-valued function defined on
$
G \times X
$
with complex entries. The indices
$
\alpha, \beta
$
take values
$
1,2,...,M
$
where
$M$
is integer or even infinite.

{}From the previous equation one easily obtains a consistency condition:
\be
c(g_{1},g_{2}\cdot x) c(g_{2},x) = c(g_{1}g_{2}, x)
\ee
i.e in the cohomology terminology \cite{V},
$c$
is a
$
(G,X,{\cal K})$-cocycle. Here
$
{\cal K} = GL(M,{\bf C})
$.

Moreover, if we have two cohomologous cocycles,
$
c^{(1)}
$
and
$
c^{(2)}
$
i.e there exists a map
$
b: X \mapsto {\cal K}
$
such that
\be
c^{(2)}(g,x) = b(g\cdot x) c^{(1)}(g,x) b(x)^{-1}
\ee
and the field
$
\phi^{(1)}
$
verifies (1) with
$
c = c^{(1)}
$,
let us define
\be
\phi^{(2)}_{\alpha}(x) = b_{\alpha\beta} \phi^{(1)}_{\beta}(x).
\ee
Then
$
\phi^{(2)}
$
verifies (1) with
$
c = c^{(2)}.
$

This argument shows that in order to classify all possible
relativistic laws of transformation of the type (1) it is
sufficient to take a representative cocycle
$c$
from every cohomology class. According to the general theory
from \cite{V}, this analysis is possible if
$G$
acts transitively on
$X$.
Then the cohomology classes are in one-one correspondence with
the equivalence classes of representation of the stability group
of some point
$
x_{0} \in X.
$

The case when
$G$
admits non-trivial multiplicators can be analysed on similar
lines. The relation (2) will be true up to some phase factor
depending on
$
g_{1}
$
and
$
g_{2}.
$

2.2 A particular important case is when
$
X = M_{N}
$
(i.e. the
$N$-dimensional Minkowski space) and
$
G = \up = H \times_{t} A
$
with
$A$
the translation group of
$
M_{N},
$
and
$
H = \ul
$.
The action of
$G$
on
$X$
is
\be
(h,a)\cdot x = h\cdot x + a.
\ee

This action is transitive and taking
$
x_{0} = 0
$
we have
$
G_{0} = \{ (h,0) \vert h \in H \} \simeq H.
$
So, the the cohomology classes of cocycles
$c$
are indexed by equivalence classes of representations of
$H$.
Let
$
h \mapsto D(h)
$
be such a representation of
$H$
in
$
{\bf C}^{M}
$;
$D$
is not necessarely unitary. Then the cocycle induced by
$D$
is simply
\be
c^{D}((h,a),x) = D(h).
\ee

So, (1) takes the usual form (in matrix notations):
\be
{\cal W}_{h,a} \phi(x) {\cal W}_{h,a}^{-1} = D(h^{-1})\phi(h\cdot x+a).
\ee
We note in closing this subsection that one usually supposes
that on
$X$
there exists a supplementary structure of causality which is
$G$-invariant and then one requires that
$
\phi_{\alpha}(x)
$
commute or anticommute with
$
\phi_{\beta}(y)
$
and
$
\phi_{\beta}(y)^{*}
$
for
$x$
and
$y$
causally separated.

2.3 Let us suppose now that, in the framework of the preceeding
subsection,
$
\phi_{\alpha}(x)
$
is a free field. Then
$
{\cal H}
$
is a Fock space
$
{\cal H} = \oplus_{n=0}^{\infty} {\cal H}^{(n)}
$
with annihilation and creation operators
$
a(f),~a^{*}(f),~~(\forall f \in  {\cal H}^{(1)})
$
\cite{RS}.

Then (7) is true for positive and negative frequency parts
$
\phi^{(\pm)}_{\alpha}(x)
$
i.e. we have:
\be
{\cal W}_{h,a} \phi^{(\pm)}(x) {\cal W}_{h,a}^{-1} =
D(h^{-1})\phi^{(\pm)}(h\cdot x+a).
\ee

We make now the {\it Weinberg anszatz}, namely:
\be
\phi^{(+)}(x) = a(f_{\alpha,x})
\ee
where
$
f_{\alpha,x}
$
are some generalized function from the one-particle Hilbert space
$
{\cal H}^{(1)}.
$

It is well known \cite{V} that
$
{\cal H}^{(1)}
$
can be exhibited in the form
\be
{\cal H}^{(1)} = \{ \Phi: \xmp \mapsto {\bf C}^{M} \vert \int_{\xmp}
\masp \vert \Phi(p)\vert^{2} < \infty \}
\ee
and the representation of
$G$
in
$
{\cal H}^{(1)}
$
of the form:
\be
\left(W_{h,a} \Phi\right)_{i}(p) = e^{i p\cdot x} C(h, h^{-1}\cdot
p)_{ij} \Phi_{j}(h^{-1}\cdot p).
\ee

Here
$C$
is a
$
(H,\xmp,{\bf C}^{M})$-cocycle;
the formulas above are known under the name of Wigner
formalism. The representation
${\cal W}$
acting in
${\cal H}$
is given by:
\be
{\cal W}_{h,a} \equiv \Gamma(W_{h,a}).
\ee

Then one knows that:
\be
{\cal W}_{h,a} a(f) {\cal W}_{h,a}^{-1} = a(W_{h,a}f).
\ee

Also one has:
\be
\left( a(f)\Phi\right)^{(n)}_{i_{1},...,i_{n}}
(p_{1},...,p_{n}) = \sqrt{n+1}
\int_{\xmp} \masp (p_{0}) \overline{f_{i_{0}}(p_{0})}
\Phi_{i_{0},...,i_{n}}(p_{0},...,p_{n}).
\ee

So, one can insert everything in (8) and obtain immediately:
\be
W_{h,a} f_{\alpha,x} = \overline{D_{\alpha\beta}(h^{-1})}
f_{\beta,h\cdot x+a}.
\ee

{}From this relation it easily follows that
$
f_{\alpha,x}
$
is of the form:
\be
f_{\alpha,x}(p) = e^{i p\cdot x} f_{\alpha}(p)
\ee
where
$
f_{\alpha}
$
verifies the relation:
\be
f_{\alpha}(h\cdot p) = \overline{D_{\alpha\beta}(h)} C(h,p)
f_{\beta}(p).
\ee

It is known \cite{V} that  the cocycle
$C$
is induced, in the sense of Mackey, by some unitary representation
$
\pi
$
of the "little group"
$
H_{p_{0}},~(p_{0} \in \xmp)
$
given by:
\be
\pi(h_{0}) \equiv C(h_{0},p_{0}),~~(\forall h_{0} \in H_{p_{0}}).
\ee

Then (17) implies that
\be
f_{\alpha}(p_{0}) = \overline{D_{\alpha\beta}(h)} \pi(h_{0})
f_{\beta}(p_{0}),~~(\forall h_{0} \in H_{p_{0}}).
\ee

If
$
f_{\alpha}(p_{0})
$
is a solution of this equation, then we have from (17) that:
\be
f_{\alpha}(p) = \overline{D_{\alpha\beta}}(\sigma(p)) C(\sigma(p),p_{0})
f_{\beta}(p_{0})
\ee
where
$
\sigma: \xmp \mapsto H
$
is a Borel cross section, i.e. we have
$
\sigma(p_{0}) = e_{H}
$
and
$
\sigma(p)\cdot p_{0} = p
$.

It follows that the key relation is (19) which says that
$
f_{\alpha,i}(p_{0})
$
is an invariant vector for the representation
$
\overline{D} \otimes \pi
$
of
$
H_{p_{0}}.
$
This is possible {\it iff} the Clebsh-Gordan series of
$
\overline{D} \otimes \pi
$
contains the trivial representation.

The same argument goes for
\be
\phi^{(-)}_{\alpha}(x) = a^{*}(g_{\alpha,x})
\ee
with the result that
$
\overline{D} \otimes \overline{\pi}
$
should also contain the trivial representation. So we have:

{\bf Theorem 1:} {\it A necessary and sufficient condition for
the existence of the free field verifying (7) where
$W$
is given by (12), is that the Clebsh-Gordan series of
$
D \otimes \pi
$
and
$
D \otimes \overline{\pi}
$
should contain the trivial representation. In this case the free
field is given by the formulas (9)+(21) where
$
f_{\alpha,x}
$
and
$
g_{\alpha,x}
$
are given by formulas of the type (20).}

{\bf Proof:} One has to show that the analysis above does not
depend on the explicit realization of
$
{\cal H}^{(1)}
$
and
$W$.
(One could exhibit for instance
$W$
in the Hilbert space fiber bundle formalism {\cite V}.)
But one knows that if
$
{\cal H'}^{(1)}
$
and
$W'$
is another representation of
$G$
induced also by
$\pi$,
then there exists an unitary operator
$
V: {\cal H'}^{(1)} \mapsto {\cal H}^{(1)}
$
intertwining
$W$
and
$W'$.
One extends
$V$
to the whole Hilbert space in a natural way, obtains a
relation of the type (15) with
$
f_{\alpha,x} \mapsto V f_{\alpha,x}
$
and the previous analysis remains unchanged.
$\Box$

{\bf Remarks}

1) Let us suppose now that
$
D^{(1)}
$
and
$
D^{(2)}
$
are two representations of
$
H_{p_{0}}
$
admitting the direct sum decomposition:
\be
D^{(i)} = \oplus_{\alpha_{i} \in I_{i}} \pi^{(\alpha_{i})},~~(i = 1,2)
\ee
for some index sets
$
I_{i}.
$
Then:
\be
D^{(1)} \otimes D^{(2)} = \sum_{\alpha_{1} \in I_{1},\alpha_{2} \in
I_{2}} \pi^{(\alpha_{1})} \otimes \pi^{(\alpha_{2})}
\ee
so
$
D^{(1)} \otimes D^{(2)}
$
contains the identity representation {\it iff}
there exists
$
\alpha_{1} \in I_{1},\alpha_{2} \in I_{2}
$
such that
$
\pi^{(\alpha_{1})} \otimes \pi^{(\alpha_{2})}
$
contains the trivial representation. Applying Schur lemma it
follows that
$
(\pi^{(\alpha_{1})})^{t} \simeq \pi^{(\alpha_{2})}
$
where
$
(\pi^{(\alpha)})^{t}_{ij}(h) \equiv \pi^{(\alpha)}_{ji}(h^{-1}).
$

In our case
$
D^{(1)} \mapsto D\vert_{H_{p_{0}}}
$
and
$
D^{(2)} \mapsto \pi, \overline{\pi}.
$
For an elementary particle,
$
\pi
$
is irreducible, so the criterion above sums up to the
requirement that
$
D\vert_{H_{p_{0}}}
$
should contain
$
\pi^{t}
$
and
$
\overline{\pi}^{t}
$
as subrepresentations.

If we want that the equations of the type (19) have an unique
solution, then these subrepresentations should appear only once.

2) If
$
N = M
$,
the preceeding remark implies that
$
D\vert_{H_{p_{0}}} \simeq \pi^{t}
$
and
$
D\vert_{H_{p_{0}}} \simeq \overline{\pi}^{t}
$
so
$
\pi
$
should be self-contragradient.

\section{Applications}

3.1 First, consider the case
$
N = 4
$
i.e. the four-dimensional Minkowski space to the purpose of
obtaining the results of Weinberg. One knows \cite{V} that it
is possible to take
$
H = SL(2,{\bf C}).
$
For particles of non-zero mass
$
m \in \R^{*},
$
and spin
$
s \in {\bf Z}/2
$
we have
$
H_{p_{0}} \simeq SU(2)
$
and
$
\pi \simeq D^{(s)}
$
(i.e. the irreducible representation of weight
$s$
of
$SU(2)$~.)
As
$
D^{(s)}
$
is self-contragradient, we are looking for a representation
$D$
of
$
SL(2,{\bf C})
$
such that
$
D\vert_{H_{p_{0}}}
$
contains
$
D^{(s)}.
$
If this subrepresentation is contained only once, then the free
field is essentially unique. Such a representation is, in
standard notations,
$
D^{(s,0)}.
$
The construction of the corresponding free field according to
the scheme from subsection 2.3 gives us the results from \cite{SW},
ch. 3-2. This is the result from \cite{W1}.

For particles of zero mass
$
m = 0
$
and helicity
$
n \in {\bf Z}/2,
$
one knows that
$
H_{p_{0}} \simeq SE(2)
$
and
$
\pi \simeq \pi_{n}
$
(see \cite{V} ch. IX-3.) So, we are looking for a
representation
$D$
of
$
SL(2,{\bf C})
$
such that
$
\overline{D}\vert_{H_{p_{0}}}
$
contains only once
$
\pi_{n}
$
and
$
\overline{\pi_{n}} \simeq \pi_{-n}.
$
Such a representation exists and it is
$
D^{(s_{1},s_{2})}
$
with
$
s_{1} - s_{2} = n.
$
The same assertion is true if one takes
$
\pi = \pi_{n} \oplus \pi_{-n}
$
as it is the case, for instance, for the photon. We have
recovered the results of \cite{W2}.

3.2 We consider now the case
$
N = 3
$
i.e
$X$
is the three-dimensional Minkowski space. An explicit
description of
$H$
and
$G$
can be found in \cite{G2}. For particles of non-zero mass
$
m \in \R^{*}
$
and any spin
$
s \in \R
$
we have
$
H_{p_{0}} \simeq \R
$
and
$
\pi \simeq \pi^{(s)}
$
(in the notations of \cite{G2}.) Then one tries to find a
representation
$D$
of
$H$
such that
$
\overline{D}\vert_{H_{p_{0}}}
$
contais
$
\pi^{(s)}
$
and
$
\overline{\pi^{(s)}}.
$
Such a representation have been exhibited in \cite{G1}, \cite{G2} and is
$
D = T^{(s/2,+)} \oplus T^{(-s/2,+)}
$
where
$
T^{(s,\pm)}
$
are representations from the discrete series. For particles of
zero-mass
$
m = 0
$
we have
$
H_{p_{0}} \simeq {\bf Z} \times \R
$
and
$
\pi \simeq \pi^{(s,t)}
$
(see \cite{G2}.) If
$
t = 0
$
then one can take
$D$
as before. We have recovered the results of \cite{G1}.

3.3 For completeness we also analyse the case
$
N = 1
$
i.e. the two-dimensional Minkowski space (se also \cite{B}.)
Explicitely,
$
G = \R \times \R^{2}
$
with the composition law
\be
(\chi_{1},a_{1})\cdot (\chi_{2},a_{2}) =
(\chi_{1} + \chi_{2},a_{1} + \Lambda(\chi_{1})a_{1})
\ee
where
\[\Lambda(\chi) = \left(\begin{array}{cc}
ch(\chi) & sh(\chi) \\ sh(\chi) & ch(\chi)
\end{array}\right). \]

So in this case,
$
H = \{(\chi,0)\vert \chi \in \R\} \simeq \R.
$
The new technical element in analysing this case is that
$G$
although simply connected, admits non-trivial multiplicators. So
we will have genuine projective representations. Using the usual
canonical identification
$
H^{2}(G,\R) \simeq H^{2}(Lie(G),\R)
$
valid for
$G$
connected and simply connected \cite{V} one discovers that
every multiplier of
$G$
is cohomologous to the (non-trivial) multiplier
$
m_{k}~~(k \in \R)
$
given by:
\be
m_{k}((\chi_{1},a_{1}),(\chi_{2},a_{2})) =
exp\left\{ i{k\over 2} <a_{1},\Lambda(\chi_{1})a_{2}>\right\},
\ee
where for any
$
a,b \in \R,~<\cdot,\cdot>
$
is the Lorentz invariant sesquilinear form:
\be
<a,b> \equiv a_{0} b_{1} -  a_{1} b_{0}.
\ee

Then we have

{\bf Theorem 2:} {\it Every
$
m_{k}$-projective
unitary irreducible representation of
$G$
is unitary equivalent to one of the following form:

(i) for
$
k \in \R^{*}, W^{k,c}~~(c \in \R)
$
acting in
$
L^{2}(\R,d\lambda)
$
according to:
\be
\left(W^{k,c}_{0,a}f\right)(\lambda) = exp\left\{
ia_{0}\left(\lambda+{k \over 2}a_{1}\right)\right\} f(\lambda+ka_{1})
\ee
\be
W^{k,c}_{\chi,0} = e^{ic\chi} e^{i\chi \overline{B_{0}}}
\ee
where
$
B_{0} \equiv -{1\over 2k} \left( k^{2} {\partial^{2}\over
\partial \lambda^{2}} + \lambda^{2}\right)
$
is defined on
$
{\cal C}^{\infty}(\R),
$
and
$
\overline{B_{0}}
$
is the (unique) self-adjoint extension of
$
B_{0}
$
to
$
L^{2}(\R,d\lambda).
$

(ii) for
$
k = 0,
$
we have:

(a)
$
W^{m,\eta}~~(m \in \R_{+} \cup \{0\}, \eta = \pm)
$
acting in
$
L^{2}(\xmp,\masp)
$
acording to:
\be
\left(W^{m,\eta}_{\chi,a}f\right)(p) = e^{ia\cdot
p}~f(\Lambda(\chi)^{-1}\cdot p).
\ee

(b)
$
W^{im}~~(m \in \R^{*})
$
acting in
$
L^{2}(\y,\beta_{m})
$
acording to:
\be
\left(W^{im}_{\chi,a}f\right)(p) = e^{ia\cdot
p}~f(\Lambda(\chi)^{-1}\cdot p).
\ee

(c)
$
W^{c}~~(c \in \R)
$
acting in
{\bf C}
according to
\be
W^{c}_{\chi,a} =  e^{ic\chi}
\ee}

{\bf Proof:} The case
$
k = 0
$
is standard and it is analysed as in \cite{V}. We concentrate
on the case
$
k \in \R^{*}.
$

Let
$
W_{\chi,a}
$
be a
$
m_{k}$-projective
unitary irreducible representation of
$G$.
We define
\be
U_{a} = W_{0,a},~~V_{\chi} = W_{\chi,0}
\ee
and we have:
\be
V_{\chi}V_{\chi'} = V_{\chi+\chi'}
\ee
\be
U_{a+a'} = exp\left\{ i{k\over 2} <a,a'>\right\} U_{a}U_{a'}
\ee
\be
V_{\chi} U_{a} V_{\chi}^{-1} = U_{\Lambda(\chi) a}.
\ee

Conversely, if
$
V_{\chi}
$
and
$
U_{a}
$
are (unitary) operators verifying (33)-(35) then:
\be
W_{\chi,a} \equiv  U_{a} V_{\chi}
\ee
is a
$
m_{k}$-projective
unitary irreducible representation of
$G$.
We first analyse (34). It is easy to connect
$
U_{a}
$
to a representation of the Weyl canonical commutation relations
(in general a reducible one.) Using the standard representation
theorem of Stone-von-Neumann \cite{V} we find out easily that
one can take
$
{\cal H} = L^{2}(\R,d\lambda,{\bf C}^{M})
$
(where for the moment
$M = 1,2...,\infty$
is undetermined) and
\be
\left(U_{a}f\right)(\lambda) = exp\left\{
ia_{0}\left(\lambda+{k \over 2}a_{1}\right)\right\} f(\lambda+ka_{1}).
\ee

Next, one analyses the degree of arbitrariness of
$
V_{\chi}.
$
If
$
V_{\chi}
$
and
$
V_{\chi}'
$
both verify (33) and (35) then
$
{\cal V}_{\chi} \equiv V_{\chi}^{-1} V_{\chi}'
$
commutes with
$
U_{a},
$
so, using lemma 6.4 from \cite{V}, one easily descovers that it
has the form:
\be
\left({\cal V}_{\chi}f\right)(\lambda) = g_{\chi} f(\lambda)
\ee
with
$
g_{\chi}
$
a
$
M \times M
$
complex valued matrix.

{}From (33) it follows that in fact
$
g_{\chi} = e^{iC\chi}
$
with
$C$
a real
$
M \times M
$
self-adjoint matrix. So:
$
V_{\chi}' =  e^{iC\chi} V_{\chi}.
$

According to Stone theorem,
$
V_{\chi} = e^{i\chi B}
$
with
$B$
an self-adjoint operator in
$
L^{2}(\R,d\lambda,{\bf C}^{M}).
$
Using infinitesimal arguments, one discovers that a possible
solution of (33)+(35) is
$
\overline{B_{0}}
$
where
$
B_{0} \equiv -{1\over 2k} \left( k^{2} {\partial^{2}\over
\partial \lambda^{2}} + \lambda^{2}\right).
$

According to the argument above it follows that the most general
$B$
is
$
B = \overline{B_{0}} + C
$
with
$C$
described previously.

Finally, one notices that, by a unitary transformation in
${\cal H}$,
one can take
$C$
to be diagonal. If
$
M > 1
$
then
$W$
given by (36) is reducible. So we must have
$
M = 1. \Box
$

We can analyse now the existence of free fields. For the case
$
k = 0
$
the analysis is elementary. In the case
$
k \not= 0
$
we proceed as follows. First, one shows that although we have a
genuine projective representation, relation (2) remains
unchanged. So, again we obtain as a possible transformation law
for the field a relation of the type (7). We need in the
following only the behviour with respect to translation i.e. the
case
$
h = e_{H}:
$
\be
{\cal W}_{e_{H},a} \phi_{\alpha}(x) {\cal W}_{e_{H},a}^{-1} =
\phi_{\alpha}(x+a)
\ee

As before, we take
$
{\cal H}^{(1)} = L^{2}(\R,d\lambda)
$
with the representation
$W$
given by (27)+(28),
$
{\cal H} = {\cal F}({\cal H}^{(1)})
$
with
$
{\cal W} = \Gamma(W)
$
and
$
\phi_{\alpha}^{(\pm)}
$
given by Weinberg anszatz (9)+(21). Then one easily finds out that
$
f_{\alpha,x}
$
verifies:
\be
f_{\alpha,x+a}(\lambda) =
exp\left\{ia_{0}\left(\lambda+{k \over 2}a_{1}\right)\right\}
f_{\alpha,x}(\lambda)
\ee
which has only the trivial solution
$
f_{\alpha,x} = 0.
$
So in this case we cannot construct free fields.

3.4 We will exhibit the flexibility of the formalism developped
above by treating the case of a configuration space
$X$
which is not the Minkowski space. We consider
\be
X \equiv \{(x,\nu) \in M_{N} \times M_{N} \vert \nu^{2} = -1\}
\ee
where
$
\nu^{2} = \nu^{2}_{0} - {\bf\nu^{2}}
$
is the Minkowski norm of
$
\nu.
$

This configuration space can be physically interpreted as
describing cones of fixed angle \cite{M}. Then
$x$
is the position of the apex and
$
\nu
$
is giving the direction of the symmetry axis. (In the "rest
frame" where
$
\nu = (0,{\bf \nu}), {\bf \nu}
$
is precisely the direction of the symmetry axis.)

The action of the Poincar\'e group (and its universal covering
group) on
$X$
is natural:
\be
(h,a)\cdot (x,\nu) = (h\cdot x+a,h\cdot \nu).
\ee

So, now the fields are operators of the type
$
\phi_{\alpha}(x,\nu)
$
and, instead of (7), we will have:
\be
{\cal W}_{h,a} \phi(x,\nu) {\cal W}_{h,a}^{-1} =  D(h^{-1})
\phi(h\cdot x+a,h\cdot\nu)
\ee

It is natural to assume, instead of (9) something like:
\be
\phi_{\alpha}^{+}(x,\nu) = a(f_{\alpha,x,\nu}).
\ee

Then the analysis of the subsection 2.3 can be followed
practically unchanged. One gets (15) with obvious modifications.
So we keep a relation of the type (16), but instead of (17) we have:
\be
\phi_{\alpha,x,\nu}(h\cdot p) = \overline{D_{\alpha\beta}(h)}
C(h,p) \phi_{\beta,x,\nu}(p).
\ee

We take here
$
p = p_{0},~\nu = \nu_{\rho} \equiv (\rho,0,...,0,\sqrt{\rho^{2}+1}),
$
(where e.g
$
\rho = \nu\cdot p/m$
in the case
$
m \not= 0),
$
and
$$
h \in H_{p_{0},\rho} \equiv \{(h_{0} \in H \vert h_{0}\cdot
p_{0} = p_{0},~h_{0}\cdot \nu_{\rho} = \nu_{\rho}\}.
$$

One obtains, instead of (19), for any
$
h_{0} \in H_{p_{0},\rho}
$
\be
f_{\alpha,\nu_{\rho}}(p_{0}) =
\pi(h_{0}) \overline{D_{\alpha\beta}(h_{0})} f_{\beta,\nu_{\rho}}(p_{0})
\ee
(for
$
N = 4~ H_{p_{0},\rho} \simeq SU(2),
$
and for
$
N = 3, H_{p_{0},\rho} \simeq {\bf Z}
$)
and one can write a theorem of the same type as theorem 1.

Let us note in closing that the construction of the free fields
based on Weinberg anszatz is extremely sensitive to the choice
of the configuration space
$X$.
For instance, let us take (instead of (42) )
$X$
to be the manifold of straight lines in the Minkowski space
$
M_{N}.
$
This space can be obtained from (41) factorizing to the
equivalence relation
$
(x,\nu) \sim (x',\nu') \Longleftrightarrow \nu = \nu',~x - x'
$
is parallel to
$
\nu.
$
We will get something of the type (45) with
$
(x,\nu) \mapsto [x,\nu]
$
(i.e. the equivalence class of
$(x,\nu)
$)
and one easily obtains
$
f_{\alpha,[x,\nu]} = 0.
$
So, in this case we cannot construct free fields.

\end{document}